\documentstyle[12pt,epsf]{article}
\newcommand{\TPG}{T_{\rm PG}}

\newcommand{\Tc}{T_{\rm c}}

\newcommand{\TtG}{T_{2{\rm G}}}

\begin{document}
\begin{center}
\LARGE {\bf Theory of Electron Differentiation, Flat Dispersion and Pseudogap Phenomena}\\
\vspace{5mm}
\large{Masatoshi Imada$^{1)}$ and
Shigeki Onoda}\\
\vspace{5mm}
\normalsize
{ Institute for Solid State Physics, University of Tokyo, 
Roppongi, Minato-ku, Tokyo 106-8666, Japan}\\
{1) e-mail address, imada@issp.u-tokyo.ac.jp}
\end{center}

\baselineskip 18pt

\begin{abstract}
Aspects of electron critical differentiation are clarified in the proximity of the Mott insulator.  The flattening of the quasiparticle dispersion appears around momenta $(\pi,0)$ and $(0,\pi)$ on square lattices and determines the criticality of the metal-insulator transition with the suppressed coherence in that momentum region of quasiparticles. Such coherence suppression at the same time causes an instability to the superconducting state if a proper incoherent process is retained.  The d-wave pairing interaction is generated from such retained processes without disturbance from the coherent single-particle excitations.  Pseudogap phenomena widely observed in the underdoped cuprates are then naturally understood from the mode-mode coupling of d-wave superconducting(dSC) fluctuations with antiferromagnetic ones.  When we 
assume the existence of a strong d-wave pairing force repulsively competing with 
antiferromagnetic(AFM) fluctuations under the formation of flat and damped 
single-particle dispersion, we reproduce 
basic properties of the pseudogap seen in the magnetic resonance, neutron 
scattering, angle resolved photoemission and tunneling measurements in the 
cuprates.

\end{abstract}
\setlength{\leftmargin}{-30mm}


\section{Introduction}

High-temperature cuprate superconductors show a variety of unusual properties in 
the normal state~\cite{RMP}.  Among others, we discuss two remarkable properties widely 
observed in the cuprates.  One is the flat dispersion in single-particle 
excitations.  Angle resolved photoemission spectra (ARPES) in Y and Bi based high-Tc 
cuprates show an unusual dispersion which is far from weak correlation 
picture~\cite{Gofron}.  The dispersion around $(\pi,0)$ and $(0,\pi)$ is extremely 
flat beyond the expectations from usual van Hove singulatrities.  The flat 
dispersion also shows rather strong damping.  The other property we discuss is the 
pseudogap phenomenon observed in the underdoped region. It is observed both in 
spin and charge
excitations in which gap structure emerges from a temperature $\TPG$ well above 
the superconducting
transition point $\Tc$.  The gap structure is observed  in various different  
probes such as NMR
relaxation time, the Knight shift, neutron scattering, tunnenling,
ARPES, specific heat, optical conductivity, and DC
resistivity.  The 
ARPES~\cite{LoeserShenDessauMarshallParkFournierKapitulnik1996,photoemission}
data have revealed that the pseudogap starts growing first in the
region around $(\pi,0)$ and $(0,\pi)$ from $T=\TPG$ much higher than $T_c$.  
Therefore, the pseudogap appears from the momentum region of the flattened  
dispersion and it is likely that the mechanism of the pseudogap formation is deeply 
influenced from the underlying flatness.   The superconducting state itself also 
shows a dominant gap structure in this flat spots, $(\pi,0)$ and $(0,\pi)$,
 due to 
the $d_{x^2-y^2}$ symmetry. In fact, the pseudogap structure above $T_c$ 
continuously merges into the $d_{x^2-y^2}$ gap below $\Tc$.  To understand the 
pairing mechanism and the origin of the high transition temperatures, a detailed 
understanding of the physics taking place in the flat dispersion region is 
required.  

\section{Flat Dispersion}

Theoretically, the emergence of the flat dispersion around $(\pi,0)$ and $(0,\pi)$ 
has also been reported in numerical simulation results rather universally in 
models for strongly correlated electrons such as the Hubbard and $t$-$J$ models on 
square lattices~\cite{Assaad98}. Furthermore, the emergence of the flat dispersion also leads to 
various unusual properties.  For example, the electronic compressibility 
critically diverges as $\kappa \propto 1/\delta$  with decreasing doping 
concentration $\delta$ while the Drude weight is unusually suppressed as $D 
\propto \delta^2$~\cite{Furukawa93,Tsune98}. In more comprehensive understanding, the emergence of the flat 
dispersion is connected with the criticality of the transition from metals to 
the Mott insulator.  All the numerical data are consistent with the hyperscaling 
relations with a large dynamical exponent $z=4$ for the metal-insulator 
transition~\cite{RMP,Imada95}.  Such large exponent opposed to the usual value $z=2$ for the 
transition to the band insulator is derived from the slower electron dynamics 
generated by the flat dispersion.  This criticality is interpreted from a strong 
proximity of the Mott insulator where strong electron correlation generates 
suppressed dynamics and coherence.  The coherence temperature (the effective Fermi 
temperature) is also scaled as $T_F\propto \delta^2$ and indicates unusual 
suppression.  

In a simple picture, the correlation effects emerge as the isotropic mass 
renormalization, where the Coulomb repulsion from other electrons makes the 
effective mass heavier.  This effect was first demonstrated by Brinkman and Rice~\cite{Brinkman}  
in the Gutzwiller approximation and refined in the dynamical mean field theory 
~\cite{DMFT}.   
In the numerical results on a square lattice, the correlation effects appear in 
more subtle way where the electrons at different momenta show different 
renormalizations.  When the Mott insulator is approached and the doping 
concentration becomes small, the mass renormalization generally becomes stronger.  
 However, once the renormalization effect gets relatively stronger in a part of 
the Fermi surface, it is further enhanced at that part in a selfconsistent fashion because the 
slower electrons become more and more sensitive to the correlation effect. This 
generates critical differentiation of the carriers depending on the portion of the Fermi surface.  
On square lattices, the stronger renormalization happens around ($\pi,0$) and ($0,\pi$). 
The trigger of this strong correlation effect concentrating near
($\pi,0$) and ($0,\pi$) is intuitively understood from the carrier
motion under the background of antiferromagnetic correlations.  As we
see a real space picture in Fig.1a, the carrier motion in the diagonal 
direction does not disturb the correlations due to the parallel spin
alignment, while the motion in horizontal and vertical directions
strongly disturbs the AFM backgrounds and the motion itself is also
disturbed as a feedback.  
Such strong coupling of charge dynamics to spin correlations causes
flattering and damping of electrons around ($\pi,0$) and ($0,\pi$),
but not around the diagonal direction
($\pm {\pi}/{2},\pm {\pi}/{2}$).  
The  anisotropic renormalization effect eventually generates a singularly flat 
dispersion on particular region of the Fermi surface, which accepts more and more
doped holes in that region due to the enhanced density of states.   
The transition to the Mott insulator is 
then governed by that flattened part, since the carriers reside
predominantly 
in the flat region.    
The criticality of the metal-insulator transition on the 
square lattice is then determined from the doped carriers around the flat spots, 
$(\pi,0)$ and $(0,\pi)$.  The hyperscaling relation becomes satisfied because such 
singular points on the momentum space governs the transition.  

We, however, should keep in mind that the relaxation time of quasiparticles 
and the damping constant of magnetic excitations do not have criticality at 
the transition point to the Mott insulator.  A general remark is that the 
relaxation time is critical only in the case of the Anderson localization 
transition and not in the case of the transition to the Mott insulator.  The DC 
transport properties and magnetic relaxation phenomena are contaminated by such 
noncritical relaxation times $\tau$ and are influenced by the carriers in the 
other portion than the flat part because the flat part has stronger damping and 
less contributes to the DC properties.  Large anisotropy of $\tau$ masks the real 
criticality and makes it difficult to see the real critical exponents in the 
$\tau$-dependent properties.  Relevant quantities to easily estimate the 
criticality is the $\tau$ independent quantities such as the Drude weight and the 
compressibility.  

Near the metal-insulator transition, critical electron 
differentiation and selective renormalization may lead to experimental 
observations as if internal degrees of freedom of the carriers such as spin and 
charge were separated because each degrees of freedom can predominantly be 
conveyed by carriers in different part of the Fermi surface.  Another possible 
effect of the electron differentiation is the appearance of several different 
relaxation times which are all originally given by a single quasiparticle 
relaxation in the isotropic Fermi liquids, but now depend on momenta of the quasiparticles.  

If the mass renormalization would happen in an isotrpic way as in the picture of 
Brinkman and Rice, the renormalization can become stronger without disturbance 
when the insulator is approached.  However, if the singularly renormalized flat 
dispersion emerges critically only in a part near the Fermi surface
but the whole band width is ratained, that flattened 
part has stronger instability due to the coupling to larger energy scale retained 
in other part of the momentum space. 
The instability can be mediated by local and incoherent carrier motion
generated from two-particle processes derived in the strong coupling
expansion~\cite{Tsunetsugu99}.  
The instability of the flat dispersion was 
studied by taking account such incoherent terms in the  Hubbard and $t$-$J$ models~\cite{Assaad982,Assaad983,Tsune98}.  The 
inclusion of the two-particle terms drives the instability of the flat part to the 
superconducting pairing and the formation of the gap structure.    The paired 
bound particles formed from two quasiparticles at the flat spots have different 
dynamics from the original quasiparticle.  In fact, when the paired singlet 
becomes the dominant carrier, the criticality changes from $z=4$ to
$z=2$, resulting in the recovery of coherence and kinetic energy gain.
It generates a strong pairing interaction from the kinetic origin.  This pairing mechanism is a consequence of suppressed single-particle coherence and electron differentiation due to strong correlations.

The instability of the flat dispersion coexisting with relatively large incoherent 
process was further studied~\cite{Tsunetsugu99,Imada001,Imada002,Kohno00}. It has turned out that promotion of the above scaling 
behavior and the 
flat dispersion offers a way to control potential instabilities.  
Even when a flat {\it band} dispersion is designed near the Fermi level by 
controlling lattice geometry and parameters, it enlarges the critical region 
under the suppression of single-particle coherence in the
proximity of the Mott insulator mentioned above.
In designed lattices and lattices with tuned lattice parameters, it was reported 
that the superconducting instability and the formation of the spin gap have been 
dramatically enhanced~\cite{Imada001}.  

\section{Theory of Pseudogap Phenomena}

As is mentioned in \S1, the pseudogap starts growing from the region
of the flat dispersion.  When the single-particle coherence is
suppressed, the system is subject to two particle instabilities.  As
clarified in \S2, the superconducting instability in fact grows.
However, the antiferromagnetic and charge order correlations are in
principle also expected to grow from other two-particle
(particle-hole) processes and may compete each other.  In particular, the antiferromagnetic long-range order is realized in the Mott insulator and its short-range correlation is well retained in the underdoped region.  Therefore, to understand how the superconducting phase appears in the underdoped region, at least competition of dSC and AFM correlations have to be treated with underlying suppressed coherence in the region of ($\pi,0$) and ($0,\pi$).  The authors have developed a framework to treat the competition by employing the mode-mode coupling theory of dSC and AFM fluctuations where these two fluctuations are treated on an equal footing~\cite{Onoda991,Onoda992}.  

It should be noted that the strong dSC pairing interaction is resulted from a
highly correlated effect with electron differentiation while critical differentiations have not been successfully reproduced from the diagrammatic approach so far.  
Then, within the framework of the mode-mode coupling theory, at the starting point, we have assumed the existence of correlation effects leading to the flattened dispersion and the $d$-wave pairing force.  The AFM and dSC fluctuations are predominantly generated by the contributions from the quasiparticle excitations in the flatteened regions ($\pi,0$) and ($0,\pi$).  These fluctuations are treated in a set of selfconsistent equations with mode couplings of dSC and AF.  From the selfconsistent solution, the pseudogap formation is well reproduced in a region of the parameter space.  The pseudogap emerges when the mode coupling between dSC and AFM is repulsive with a severe competition and dSC eventually dominates at low temperatures.  Such competition suppresses $T_c$, while above $T_c$ it produces a region where pairing fluctuations are large.  
This region at  $T_{PG}>T>T_c$ shows suppression of $1/T_1T$ and the pseudogap formation around ($\pi,0$) and ($0,\pi$) in $A(k,\omega)$.  
These reproduce the basic feature of the pseudogap phenomena experimentally observed in the underdoped cuprates.  
The pseudogap formation is identified as coming from the preformed pair fluctuations.
In Fig.2, we show results obtained for the parameter values of the underdoped cuprates such as Bi2212 with $T_c\simeq$83K. The pseudogap formation in Im$G$ for {\bf k} close to ($\pi,0$) (actually =($\pi, 3\pi/64$)) is plotted at several choices of temperatures in Fig.2(a).  
The momentum dependence shows that the pseudogap formation starts around ($\pi,0$) from higher temperatures and the formation temperature becomes lower with increasing distance from ($\pi,0$) as seen in Fig.2(b).  
When we see ${\rm Im}G(k,\omega)$ around ($\pi,0$) it shows ``fill-in'' crossover with increasing temperature where the gap amplitude appears to be retained while the spectral weight inside the gap is gradually filled with the increase in temperature.  
All of the above reproduce the experimental observations.  

We, however, note a richer structure of the gap formation observed in the transversal NMR relaxation time $T_{2G}$ and the neutron resonance peak.  
One puzzling experimental observation is that the pseudogap
structure appears in $1/T_1T$~\cite{Yasuoka,Y124NMR,Bi2212NMR,Hggap,Hg1223NMR},
while in many cases $1/\TtG$, which measures ${\rm Re} \chi(Q,\omega=0)$ at $Q=(\pi,\pi)$, continuously increases with the decrease
in temperature with no indication of the pseudogap.  In addition, the so called
resonance peak appears in the neutron scattering experiments~\cite{Fong}.  A
resonance peak sharply grows  at a finite frequency below
$\Tc$ with some indications even at $T_c<T<T_{\rm PG}$.  This peak frequency 
$\omega^\ast$ decreases with lowering
doping concentration implying a direct and continuous 
evolution into the AFM Bragg peak in the undoped compounds.  The neutron and 
$T_{\rm 2G}$ data support the idea that the AFM fluctuations are suppressed
around $\omega=0$ but transferred to a nonzero
frequency below $\TPG$.  

To understand these features, a detailed consideration on damping of the magnetic excitations is required.  
With the increase in the pairing correlation length $\xi_d$, the pseudogap in $A(k,\omega)$ is developed.  
Since the damping is mainly from the overdamped Stoner excitations, the gap formation in $A(k,\omega)$ contributes not only to suppress growth of AFM correlation length $\xi_{\sigma}$ but also to reduce the magnetic damping because, inside the domain of the $d$-wave order, the antiferromagnetic excitations are less scattered due to the absence of low-energy quasiparticle around ($\pi,0$).  
If the quasiparticle damping is originally large around ($\pi,0$), the damping $\gamma$ can be reduced dramatically upon the pseudogap formation.  
Under this circumstance, our calculated result reproduces the resonance peak and the increase in $1/T_{2G}$ with lowering temperature at $T>T_c$ in agreement with the experimental observations in YBa$_2$Cu$_3$O$_{6.63}$, YBa$_2$Cu$_4$O$_8$  and some other underdoped compounds~\cite{Onoda991,Onoda992}.  

A subtlety arises when the damping around ($\pi/2,\pi/2$) starts
contributing.  
  This is particularly true  under the pseudogap formation.   If contributions from the 
$(\pi/2,\pi/2)$ region would be absent, the damping of the magnetic excitation 
would be strongly reduced when the pseudogap is formed around $(\pi,0)$ as we mentioned above.  
However, under the pseudogap formation, the damping can be determined by the Stoner 
continuum generated from  the  $(\pi/2,\pi/2)$ region and can remain overdamped.  
This process is in fact important if the quasiparticle damping around the  
$(\pi/2,\pi/2)$ region is large as in the case of La 214 compounds~\cite{LSCOARPES}.    The formation of the 
pseudogap itself is a rather universal consequence of the strong coupling 
superconductors.  However, the actual behavior may depend on this 
damping. If  the damping generated by the  
$(\pi/2,\pi/2)$ region is large, it sensitively destroy the resonance peak structure observed 
in the neutron scattering experimental results.  

\section{Summary and Discussion} \label{SECTION_summary}

Electron critical differentiation is a typical property of the
proximity of the Mott insulator.  The flattening of the quasiparticle
dispersion appears around momenta $(\pi,0)$ and $(0,\pi)$ on square
lattices and determines the criticality of the metal-insulator
transition with the suppressed coherence in that momentum region of
quasiparticles. Such coherence suppression at the same time causes an
instability to the superconducting state if a proper incoherent
process is retained.  The d-wave pairing interaction is generated from
such retained microscopic process in the strong coupling expansion without disturbance from the coherent single-particle excitations.  By assuming the d-wave attractive channel and the presence of strongly 
renormalized flat quasi-particle dispersion around the $(\pi,0)$ region,  we have 
considered the mode-mode coupling theory for the AFM and $d$SC fluctuations. 
The pseudogap in the high-$\Tc$ cuprates is reproduced as the
region with enhanced $d$SC correlations and is consistently explained from
precursor effects for the superconductivity.  The existence of the flat 
region plays a role to suppress the effective Fermi temperature $E_F$.  This 
suppressed $E_F$ and relatively large pairing interaction  both drive 
the system to the strong coupling region thereby leads to the pseudogap formation. 
  The pseudogap formation is also enhanced by the AFM fluctuations repulsively 
coupled with dSC fluctuations.

  The work was supported by ``Research for the 
Future" Program from
the Japan Society for the Promotion of Science under the grant number
JSPS-RFTF97P01103.

\end{document}